# Nanomechanical topological insulators with an auxiliary orbital degree of freedom

Jingwen Ma[†], Xiang Xi[†], Yuan Li, and Xiankai Sun[*]

*Department of Electronic Engineering, The Chinese University of Hong Kong, Shatin, New Territories, Hong Kong*

[†]*These authors contributed equally to this work*

[*]*Corresponding author:* xksun@cuhk.edu.hk

**Discrete degrees of freedom, such as spin and orbital, can provide intriguing strategies to manipulate electrons, photons, and phonons. With a spin degree of freedom, topological insulators have stimulated intense interests in condensed-matter physics[1], optics[2,3], acoustics[4], and mechanics[5-7]. However, orbital as another fundamental attribute in crystals has seldom been investigated in topological insulators. Here, we invent a new type of topological insulators with an auxiliary orbital degree of freedom on a nanomechanical platform. We experimentally realized nanomechanical topological insulators where the orbital can arbitrarily be manipulated by the crystal. Harnessing this unique feature, we demonstrated adiabatic transition between distinct topological edge states, which is a crucial functionality for complicated systems that involve distinct topological edge channels. Beyond the one-dimensional edge states, we further constructed zero-dimensional Dirac-vortex states. Our results have unveiled unprecedented strategies to manipulate topological phase transitions and to study topological phases of matter on an integrated platform.**

Topological insulators are a new phase of matter that behaves as an insulator in its interior but supports backscattering-immune conduction of electrons on its surface[8]. Originating from condensed-matter physics, similar concepts have entered the realms of optics[9,10], acoustics[11-13], and mechanics[14-18], and have overturned some of the traditional views on wave propagation and manipulation. Topological insulators can be realized by resorting to a spin[1-7] or valley-pseudospin[19-25] degree of freedom. In addition to the spin or valley-pseudospin, orbital is another fundamental attribute of Bloch waves which describes the spatial distribution of wave functions inside the unit cell of crystals. By harnessing the orbital degree of freedom one may realize novel functionalities and exotic topological states beyond backscattering-immune transport along edges, which are desperately desired for large-scale topological integrated circuits. For example, adiabatic transitions between different topological edge channels is a crucial functionality, which remains elusive in the conventional quantum spin Hall (QSH) and quantum valley Hall (QVH) topological insulators. Another example is the zero-dimensional topological bound states, which are robust against local perturbations and may serve as high-quality resonators in topological integrated photonic and phononic circuits. However, the orbitals of electrons in condensed matter have limited options due to constraint of materials and lack of controllability over the Coulomb force of atomic nucleus. In the photonic and phononic domains, manipulation of orbitals is also challenging because implementation of topological insulators requires careful considerations of crystalline symmetry and thus imposes strict limitations on the orbitals.

Here, we invented and experimentally demonstrated a new type of topological insulators with an auxiliary orbital degree of freedom based on a 2D nanomechanical crystal (Fig. 1a). The devices were fabricated on a silicon-nitride-on-oxide wafer by first etching small holes in the silicon nitride layer and then partially removing the underlying silicon oxide. Positions of the etched holes were carefully engineered such that the suspended silicon nitride membranes formed a hexagonal lattice with a period of $|a_1| = |a_2| = 21$ μm, where $a_1$ and $a_2$ are the basis vectors (Fig. 1a). Each unit cell (white dashed line in Fig. 1b) contains six nanomechanical membranes, which can be classified into two groups (labeled with different colors in Fig. 1b) due to the $C_3$ rotational symmetry. The geometries of these two groups of membranes are determined by the relative positions of the etched holes $(r_1, r_2) = (r_0 - \delta_i + \delta_t, r_0 - \delta_i - \delta_t/2)$ and $(r_3, r_4) = (r_0 + \delta_i + \delta_t, r_0 + \delta_i - \delta_t/2)$ with $r_0 = 3.34$ μm. Nonzero values of $\delta_i$ and $\delta_t$ break the $C_2$ inversion symmetry and $T_P$ translational symmetry along the vector $P$ (white arrow in Fig. 1b), respectively. The QSH insulators demonstrated previously[2-7] are mainly based on breaking the $T_P$ translational symmetry, which limits the orbitals of phonons to conventional $p$- and $d$-types. Here we explored a more generic type of topological insulators defined in a 2D parameter space $(\delta_i, \delta_t)$ (Fig. 1c), where both the $C_2$ inversion symmetry and $T_P$ translational symmetry are broken.

The 2D parameter space in Fig. 1c can be defined with the polar coordinates $(\delta_0, \theta)$ according to the relation $(\delta_i, \delta_t) = \delta_0 \cdot (\cos\theta, \sin\theta)$. By solving the Euler–Bernoulli equation at the $\Gamma$ point of the first Brillouin zone, we found that there always exist two pairs of degenerate eigenstates $\left|\psi_{n,\pm}^{(\theta)}\right\rangle$ ($n = 1, 2$) satisfying the relation $\left|\psi_{n,+}^{(\theta)}\right\rangle = K\left|\psi_{n,-}^{(\theta)}\right\rangle$, where $K$ is the complex conjugate operator. These states are protected by a kind of pseudo-time-reversal symmetry $T(\theta)$ and can be theoretically decomposed as

$$\left|\psi_{n,\pm}^{(\theta)}\right\rangle = |u_n(\theta)\rangle \otimes |s_{n,\pm}\rangle \quad (n = 1, 2), \tag{1}$$

where the orbital states $|u_n(\theta)\rangle$ and spin states $|s_{n,\pm}\rangle$ describe the spatial amplitude distribution and chirality of the eigenstates $\left|\psi_{n,\pm}^{(\theta)}\right\rangle$ ($n = 1, 2$) respectively (see Supplementary Information).

Fixing the parameter $\delta_0 = 300$ nm, we calculated the modal profiles of $\left|\psi_{n,\pm}^{(\theta)}\right\rangle$ ($n$ = 1, 2) at the $\Gamma$ point and found that their orbital states $|u_n(\theta)\rangle$ ($n$ = 1, 2) are uniquely determined by $\theta$ (Fig. 1c). The orbitals of the upper-band states $|u_2(\theta = \theta_0)\rangle$ (lower-band states $|u_1(\theta = \theta_0)\rangle$) are the same as those of the lower-band states $|u_1(\theta = \theta_0 + \pi)\rangle$ (upper-band states $|u_2(\theta = \theta_0 + \pi)\rangle$), indicating that the energy bands are always inverted for $\theta = \theta_0$ and $\theta = \theta_0 + \pi$. We further defined and calculated the spin Chern number for crystals with parameters ($\delta_0 \neq 0$, $\theta = \theta_0$) and ($\delta_0 \neq 0$, $\theta = \theta_0 + \pi$), and confirmed that they belong to distinct topological phases (see Supplementary Information). Interestingly, the eigenfrequencies $\omega_{n,\pm}^{(\theta)}$ ($n$ = 1, 2) at the $\Gamma$ point exhibit a doubly degenerate cone-like dispersion relation near the origin of the 2D parameter space (Fig. 1d), indicating that variation of $\theta$ has negligible influence on the degeneracy of $\left|\psi_{n,\pm}^{(\theta)}\right\rangle$ ($n$ = 1, 2). In other words, the proposed nanomechanical crystals support phononic QSH phases with an auxiliary orbital degree of freedom $|u_n(\theta)\rangle$ ($n$ = 1, 2) determined by the parameter $\theta$. The QSH topological insulators demonstrated previously are the special cases with $\theta = \pm\pi/2$. It is also interesting to note that the well-known QVH topological insulators[19-25] are closely related to the special cases with $\theta = 0$ and $\pi$.

Without loss of generality, we investigated the topological edge states at the domain wall between two nanomechanical crystals with parameters ($\delta_0 = 300$ nm, $\theta = \theta_0 = \pi/4$) and ($\delta_0 = 300$ nm, $\theta = \theta_0 + \pi = 5\pi/4$). Figure 2a shows the detailed geometry of the topological domain wall. The corresponding simulated energy band diagram in Fig. 2b confirms the presence of two topological edge states (purple dots) with an almost linear dispersion relation inside the bulk bandgap from 47.0 to 50.0 MHz (gray shaded region). The experimentally measured energy band diagram in Fig. 2c agrees well with the simulated results. It should be noted that the structure near the interfacial domain wall can slightly break the $\mathbf{T}(\theta)$ symmetry, and consequently the two counterpropagating edge states are slightly coupled to each other to form two standing-wave states at the $\Gamma$ point. The simulated modal profiles of these two standing-wave edge states are shown in Fig. 2d, e. The spatial distributions of the elastic waves were experimentally measured at the frequencies of 48.46 MHz (Fig. 2f) and 48.87 MHz (Fig. 2g). Both the simulated and measured modal profiles confirm that the edge states are highly localized at the topological domain wall. In addition to $\theta_0 = \pi/4$, the energy band diagrams of different topological edge states with other $\theta_0$ values were also experimentally demonstrated (see Supplementary Information).

Next, we characterized the propagation properties of the topological edge state along a Z-shaped domain wall (Fig. 2h). The nanomechanical crystals in the upper and lower regions (labeled with different colors in Fig. 2h) adopted the same parameters as Fig. 2a. We measured the amplitude distribution of the elastic waves at the frequency of 48.06 MHz (Fig. 2i). Figure 2j shows the experimental amplitude spectra of elastic waves measured at selected spots near (points A–C in Fig. 2h) or far away from (point D in Fig. 2h) the topological domain wall. The amplitude measured at points A–C is orders of magnitude stronger than that at point D in the bulk bandgap frequency range of 47.0–50.0 MHz (gray shaded region in Fig. 2j). We have also observed the backscattering-immune propagation of the topological edge state in the temporal domain (see Supplementary Information and Movie I). These experimental results confirm the presence of topological edge states with an auxiliary orbital degree of freedom and their topological robustness against the sharp bends.

Then, we experimentally demonstrated adiabatic transition between distinct topological edge states by continuously varying the orbital states $|u_n(\theta)\rangle$ ($n = 1, 2$) along the propagating direction. Figure 3b shows our fabricated device containing the nanomechanical crystals with $\theta = \theta_0(x)$ in region I and $\theta = \theta_0(x) + \pi$ in region II, where $\theta_0(x)$ increases linearly from $-\pi$ to 0 within a propagation distance $L = 1.071$ mm along the $x$ direction (Fig. 3a). Figure 3c shows the measured spatial distribution of elastic waves at the frequency of 48.35 MHz. Figure 3d shows the experimental amplitude spectra of elastic waves measured at selected spots (points A–C in Fig. 3b). In the bulk bandgap frequency range of 47.0–50.0 MHz (gray shaded region), the amplitude at points A and B is orders of magnitude weaker than that at point C. It is clear that the topological edge state between region I ($\theta = -\pi$) and region II ($\theta = 0$) could be transformed smoothly into another type of topological edge state between region I ($\theta = 0$) and region II ($\theta = \pi$).

We also experimentally demonstrated that such an adiabatic topological phase transition cannot be realized with fixed orbitals. Taking the topological edge states with fixed orbitals $|u_n(\theta = 0, \pi)\rangle$ ($n = 1, 2$) as an example, abrupt change of $\theta$ from $-\pi$ to 0 in region I and from 0 to $\pi$ in region II yields unavoidable crossing of a point with $\delta_0 = 0$ (blue point in Fig. 3e). This point introduces an additional topological domain wall (blue dashed line in Fig. 3f) orthogonal to the original domain wall (black solid line in Fig. 3f). Figure 3g shows the measured spatial distribution of elastic waves at the frequency of 48.35 MHz, which confirms that a large portion of the elastic waves was scattered into the undesired domain wall (blue dashed line in Fig. 3f). Figure 3h shows the experimental spectra measured at points A–C in Fig. 3f. These experimental results suggest that abrupt transition between distinct topological edge states suffered from strong scattering.

Finally, by using the auxiliary orbital degree of freedom we constructed zero-dimensional Dirac vortex states, which are the nanomechanical analog of the well-known Majorana bound states in superconductor electronic systems[26]. We designed and fabricated a device as shown in Fig. 4a, with the detailed structure near the center shown in Fig. 4b. The nanomechanical crystals have parameters $\delta_0(\mathbf{r}) = \delta_{max} \cdot \tan(R/R_0)$ and $\theta(\mathbf{r}) = n \cdot \varphi$, both of which are a function of the position **r** = $R \cdot (\cos\varphi, \sin\varphi)$. The parameters adopted in our experiment were $R_0 = a/10$, $\delta_{max} = 300$ nm, and $n = 1$. The detailed structure near the center of the vortex (white hexagon in Fig. 4b) indicates that the nanomechanical crystals are mirror-symmetric with respect to the axis of $\varphi = \pi/2$. Figure 4c shows the simulated modal profile of the Dirac vortex state. Figure 4d shows the experimental amplitude spectra of elastic waves measured at the white star in Fig. 4b. There exist two resonances in the bulk bandgap region (gray shaded region in Fig. 4d) with ultrahigh quality factors of ~$1.7 \times 10^4$ and ~$2.2 \times 10^4$. The amplitude distributions of the elastic waves measured at the two resonant frequencies (Fig. 4e, f) show that they resulted from hybridization of the Dirac vortex state and a defect mode induced by the electrode for actuating the elastic waves. The measured amplitude distributions of the elastic waves in Fig. 4e, f agree well with the simulated modal profile in Fig. 4c. It should be noted that the Dirac vortex states can also be realized by using a Kekulé distortion scheme[27]. Actually, these two different strategies are implemented in orthogonal 2D subspaces of an extended 3D parameter space (see Supplementary Information). When combining the orbital degree of freedom with the Kekulé distortion, it is possible to construct new topological phases, e.g., QSH insulators in four dimensions[28].

In conclusion, we have theoretically proposed and experimentally demonstrated a novel type of topological insulators with an auxiliary orbital degree of freedom. Exploiting this additional degree of freedom, we realized adiabatic topological transition between distinct edge states without closing the energy bandgap. This functionality can be used for mode conversion between distinct topological states, which enables the development of large-scale topological circuits dealing with various topological phases such as QSH and QVH states. We further realized the nanomechanical Dirac vortex states by azimuthally varying the orbital polarization of the crystals. Operating in the very high frequency regime (~50 MHz) with an ultrahigh quality factor (~$2.2 \times 10^4$), the demonstrated Dirac vortex states are topologically robust and can be used for various sensing and signal-processing applications in nanomechanical systems. Considering the large mechanical nonlinearity of the nanomechanical crystals[29], our results can be further extended to investigate the nonlinear interactions between topological phonons, which may lead to phonon lasing[30] from the topological Dirac vortex states. Although our experiment was conducted on a nanomechanical platform, the concept of exploiting the orbital degree of freedom can readily be extended and applied to other areas, such as photonics and acoustics, for extensive research of topological phase transitions and topological phases of matter.

# References

1. Bernevig, B. A., Hughes, T. L., Zhang, S.-C. Quantum spin Hall effect and topological phase transition in HgTe quantum wells. *Science* **314**, 1757–1761 (2006).
2. Wu, L.-H., Hu, X. Scheme for achieving a topological photonic crystal by using dielectric material. *Phys. Rev. Lett.* **114**, 223901 (2015).
3. Barik, S., Karasahin, A., Flower, C., Cai, T., Miyake, H., DeGottardi, W.*, et al.* A topological quantum optics interface. *Science* **359**, 666–668 (2018).
4. He, C., Ni, X., Ge, H., Sun, X.-C., Chen, Y.-B., Lu, M.-H.*, et al.* Acoustic topological insulator and robust one-way sound transport. *Nat. Phys.* **12**, 1124–1129 (2016).
5. Yu, S.-Y., He, C., Wang, Z., Liu, F.-K., Sun, X.-C., Li, Z.*, et al.* Elastic pseudospin transport for integratable topological phononic circuits. *Nat. Commun.* **9**, 3072 (2018).
6. Cha, J., Kim, K. W., Daraio, C. Experimental realization of on-chip topological nanoelectromechanical metamaterials. *Nature* **564**, 229–233 (2018).
7. Süsstrunk, R., Huber, S. D. Observation of phononic helical edge states in a mechanical topological insulator. *Science* **349**, 47–50 (2015).
8. Hasan, M. Z., Kane, C. L. Colloquium: Topological insulators. *Rev. Mod. Phys.* **82**, 3045–3067 (2010).
9. Khanikaev, A. B., Shvets, G. Two-dimensional topological photonics. *Nat. Photonics* **11**, 763–773 (2017).
10. Ozawa, T., Price, H. M., Amo, A., Goldman, N., Hafezi, M., Lu, L.*, et al.* Topological photonics. *Rev. Mod. Phys.* **91**, 015006 (2019).
11. Yang, Z., Gao, F., Shi, X., Lin, X., Gao, Z., Chong, Y.*, et al.* Topological acoustics. *Phys. Rev. Lett.* **114**, 114301 (2015).
12. He, H., Qiu, C., Ye, L., Cai, X., Fan, X., Ke, M.*, et al.* Topological negative refraction of surface acoustic waves in a Weyl phononic crystal. *Nature* **560**, 61–64 (2018).
13. Zhang, X., Xiao, M., Cheng, Y., Lu, M.-H., Christensen, J. Topological sound. *Commun. Phys.* **1**, 97 (2018).
14. Ma, G., Xiao, M., Chan, C. T. Topological phases in acoustic and mechanical systems. *Nat. Rev. Phys.* **1**, 281–294 (2019).
15. Serra-Garcia, M., Peri, V., Süsstrunk, R., Bilal, O. R., Larsen, T., Villanueva, L. G.*, et al.* Observation of a phononic quadrupole topological insulator. *Nature* **555**, 342–345 (2018).

16. Peano, V., Brendel, C., Schmidt, M., Marquardt, F. Topological phases of sound and light. *Phys. Rev. X* **5**, 031011 (2015).
17. Brendel, C., Peano, V., Painter, O. J., Marquardt, F. Pseudomagnetic fields for sound at the nanoscale. *Proc. Natl. Acad. Sci. U. S. A.* **114**, E3390–E3395 (2017).
18. Foehr, A., Bilal, O. R., Huber, S. D., Daraio, C. Spiral-based phononic plates: From wave beaming to topological insulators. *Phys. Rev. Lett.* **120**, 205501 (2018).
19. Ju, L., Shi, Z., Nair, N., Lv, Y., Jin, C., Velasco Jr, J.*, et al.* Topological valley transport at bilayer graphene domain walls. *Nature* **520**, 650–655 (2015).
20. Shalaev, M. I., Walasik, W., Tsukernik, A., Xu, Y., Litchinitser, N. M. Robust topologically protected transport in photonic crystals at telecommunication wavelengths. *Nat. Nanotechnol.* **14**, 31–34 (2019).
21. Ma, J., Xi, X., Sun, X. Topological photonic integrated circuits based on valley kink states. *Laser Photon. Rev.* **13**, 1900087 (2019).
22. Lu, J., Qiu, C., Ye, L., Fan, X., Ke, M., Zhang, F.*, et al.* Observation of topological valley transport of sound in sonic crystals. *Nat. Phys.* **13**, 369–374 (2017).
23. Zhang, Z., Tian, Y., Wang, Y., Gao, S., Cheng, Y., Liu, X.*, et al.* Directional acoustic antennas based on valley-Hall topological insulators. *Adv. Mater.* **30**, 1803229 (2018).
24. Yan, M., Lu, J., Li, F., Deng, W., Huang, X., Ma, J.*, et al.* On-chip valley topological materials for elastic wave manipulation. *Nat. Mater.* **17**, 993–998 (2018).
25. Liu, T.-W., Semperlotti, F. Tunable acoustic valley-Hall edge states in reconfigurable phononic elastic waveguides. *Phys. Rev. Appl.* **9**, 014001 (2018).
26. Jackiw, R., Rossi, P. Zero modes of the vortex-fermion system. *Nucl. Phys. B* **190**, 681–691 (1981).
27. Gao, P., Torrent, D., Cervera, F., San-Jose, P., Sánchez-Dehesa, J., Christensen, J. Majorana-like zero modes in Kekulé distorted sonic lattices. *Phys. Rev. Lett.* **123**, 196601 (2019).
28. Zilberberg, O., Huang, S., Guglielmon, J., Wang, M., Chen, K. P., Kraus, Y. E.*, et al.* Photonic topological boundary pumping as a probe of 4D quantum Hall physics. *Nature* **553**, 59–62 (2018).
29. Kurosu, M., Hatanaka, D., Onomitsu, K., Yamaguchi, H. On-chip temporal focusing of elastic waves in a phononic crystal waveguide. *Nat. Commun.* **9**, 1331 (2018).

30. Mahboob, I., Nishiguchi, K., Fujiwara, A., Yamaguchi, H. Phonon lasing in an electromechanical resonator. *Phys. Rev. Lett.* **110**, 127202 (2013).

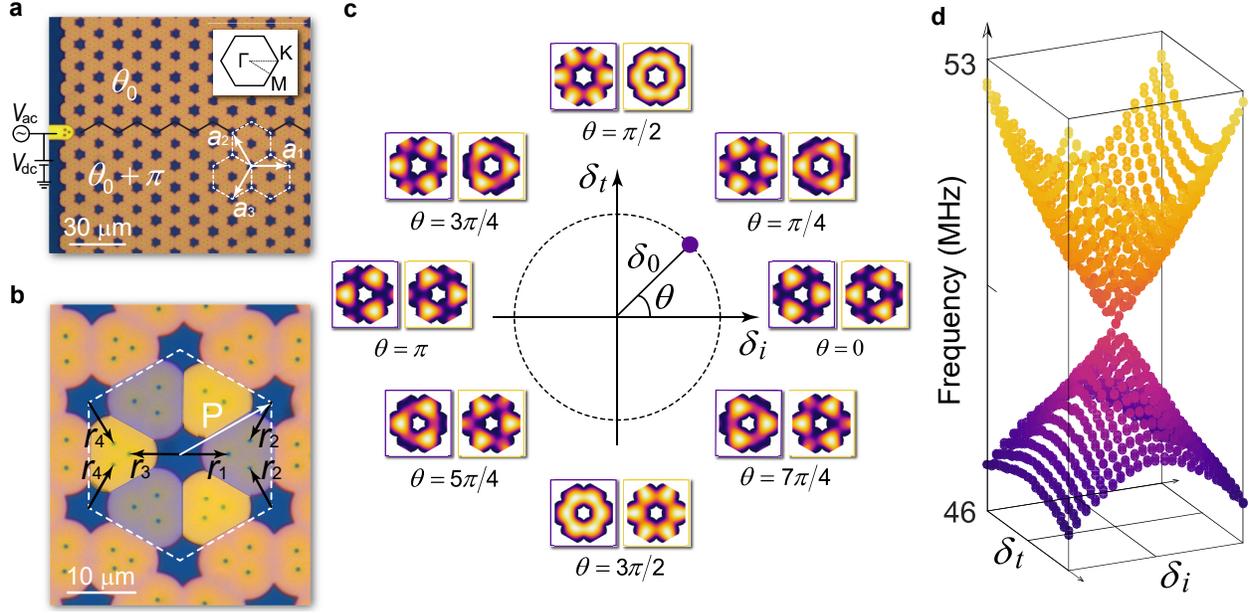

**Fig. 1 | Topological nanomechanical crystals with an auxiliary orbital degree of freedom. a**, Optical microscope image of the 2D nanomechanical crystal. The suspended silicon nitride membranes are marked in orange, and the unsuspended parts of silicon nitride are marked in dark blue. The suspended silicon nitride membranes form a 2D hexagonal lattice with basis vectors $a_1$ and $a_2$. The inset shows the first Brillouin zone. **b**, Zoomed-in optical microscope image of the nanomechanical crystal. The patterned small holes in the silicon nitride layer allowed penetration of buffered oxide etchant into the underlying oxide and thus enabled release of the nanomechanical membranes from the substrate. As indicated by the white dashed hexagon, each unit cell contains two groups of membranes labeled with different colors, whose geometries are determined by the positions of the etched holes $(r_1, r_2) = (r_0 - \delta_i + \delta_t, r_0 - \delta_i - \delta_t/2)$ and $(r_3, r_4) = (r_0 + \delta_i + \delta_t, r_0 + \delta_i - \delta_t/2)$. **c**, 2D parameter space defined by the geometric parameters $(\delta_i, \delta_t)$. The modal profiles of the lower-band states $\left|\psi_{1,\pm}^{(\theta)}\right\rangle$ and upper-band states $\left|\psi_{2,\pm}^{(\theta)}\right\rangle$ are shown in purple and orange boxes, respectively. **d**, Simulated eigenfrequencies $\omega_{n,\pm}^{(\theta)}$ ($n = 1, 2$) at the $\Gamma$ point of the nanomechanical crystals in the 2D parameter space.

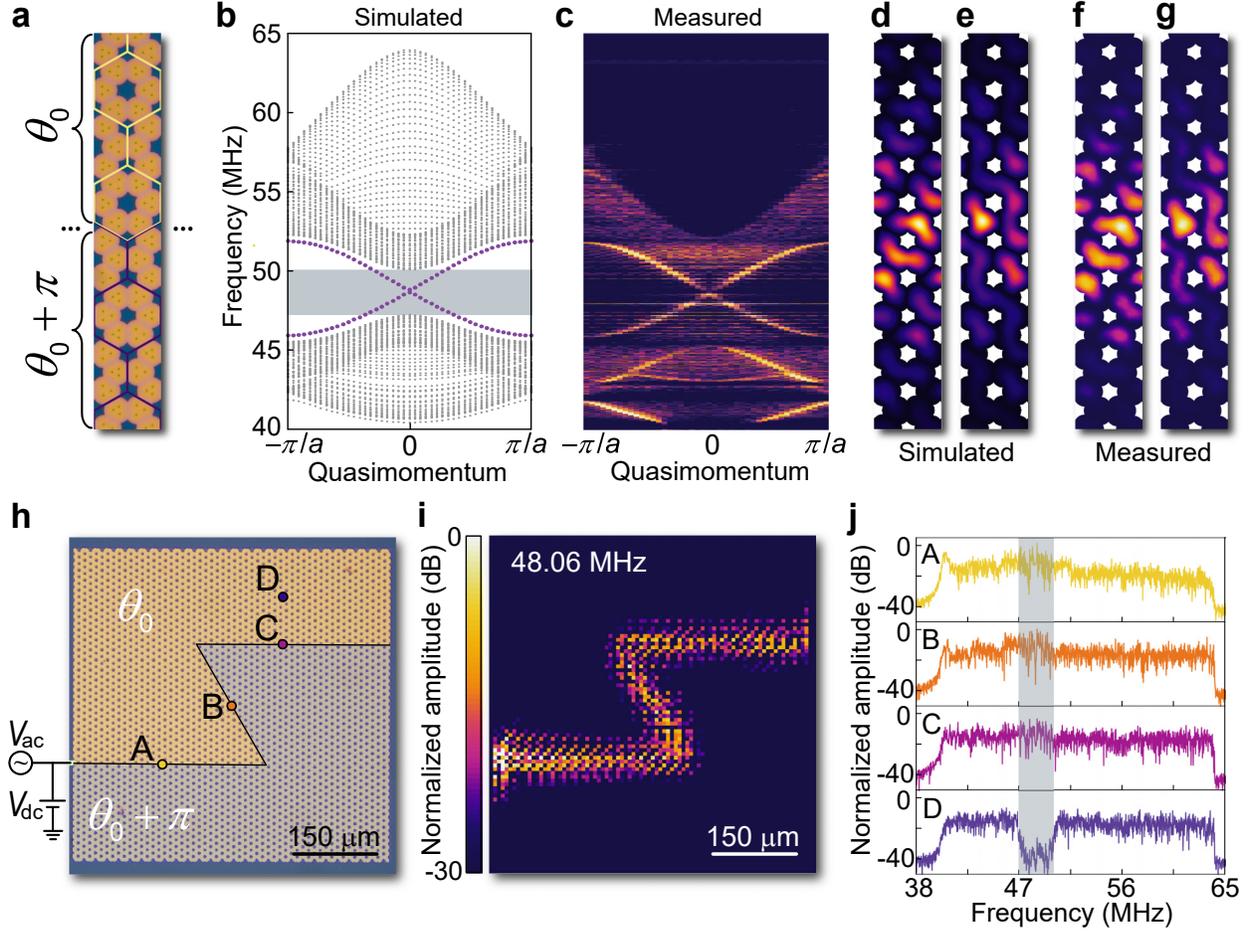

**Fig. 2 | Topological edge states with an auxiliary orbital degree of freedom. a**, Optical microscope image of the topological domain wall. The structural parameters are ($\delta_0$ = 300 nm, $\theta = \theta_0 = \pi/4$) above the domain wall and ($\delta_0$ = 300 nm, $\theta = \theta_0 + \pi = 5\pi/4$) below the domain wall. **b**, Simulated energy band diagram of the structure in **a**. The two topological edge states (purple dots) cross the entire bulk bandgap (gray shaded region). **c**, Experimental energy band diagram measured near the domain wall. **d**, **e**, Simulated modal profiles of the topological edge states at the $\Gamma$ point. **f**, **g**, Measured spatial distributions of elastic waves at the frequencies of 48.46 MHz (**f**) and 48.87 MHz (**g**). **h**, Optical microscope image of a fabricated device containing a Z-shaped topological domain wall. The flexural motions of the membranes were actuated electrocapacitively by a combination of constant voltage $V_{dc}$ and alternating voltage $V_{ac}$ applied to the excitation electrode. **i**, Measured spatial distribution of elastic waves at the frequency of 48.06 MHz. **j**, Mechanical amplitude spectra measured at spots near (points A–C in **h**) and far away from (point D in **h**) the topological domain wall in **h**. The gray shaded region indicates the bulk bandgap.

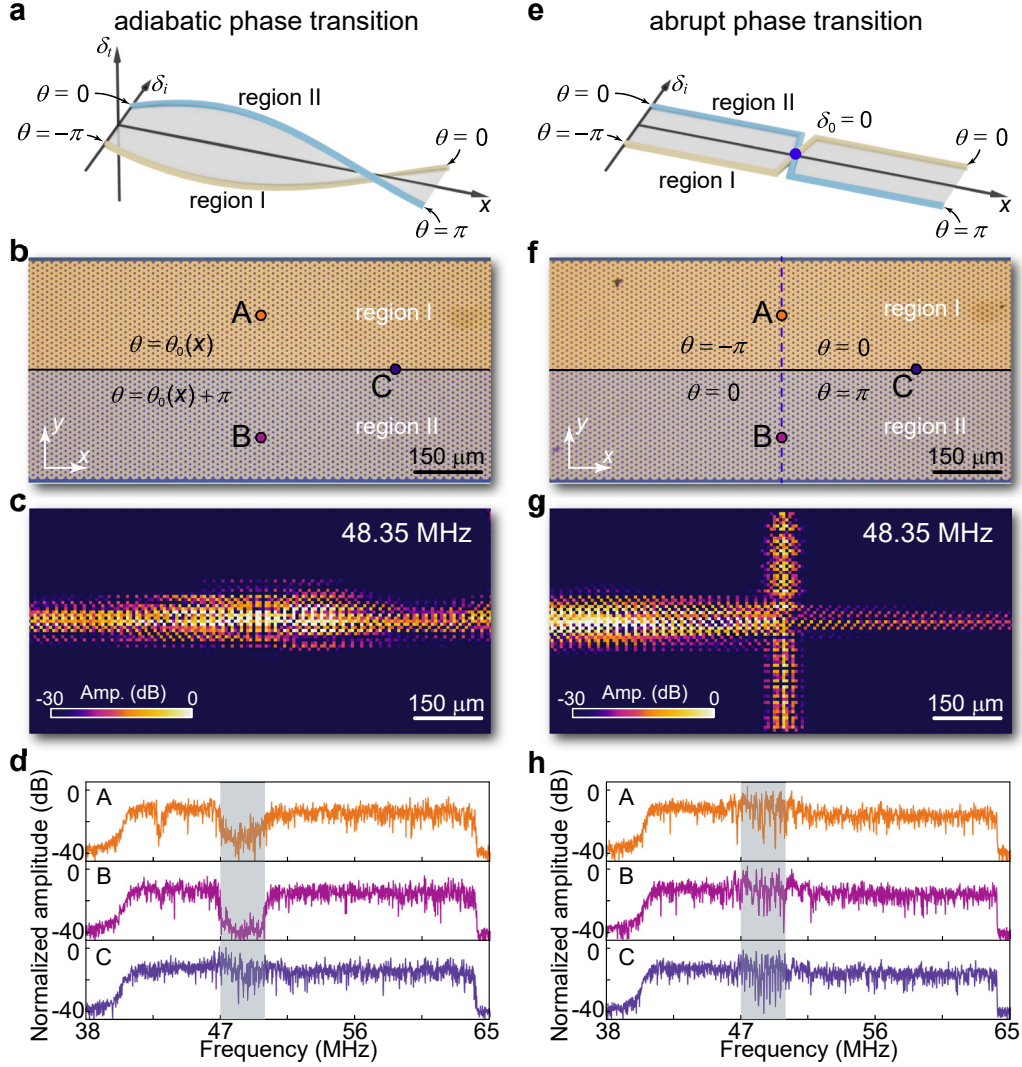

**Fig. 3 | Adiabatic and abrupt phase transitions between different topological edge states. a**, Concept of adiabatic phase transition between two different topological edge states with the auxiliary orbital degree of freedom. By fixing $\delta_0 = 300$ nm and varying $\theta$ gradually along the propagation direction (+$x$ direction), one can realize adiabatic phase transition between different topological edge states without closing the energy bandgap. **b**, Optical microscope image of the device for demonstrating adiabatic phase transition of the topological edge states. The nanomechanical crystals in regions I and II adopt the same structural parameters described in **a**. **c**, Measured spatial distribution of elastic waves at the frequency of 48.35 MHz for the device in **b**. **d**, Mechanical amplitude spectra measured at points A–C in the device in **b**. **e**, Concept of abrupt phase transition between two different topological edge states with fixed orbitals. Change of $\theta$ from $-\pi$ to 0 in region I and from 0 to $\pi$ in region II yields unavoidable crossing of an abrupt phase transition point with $\delta_0 = 0$ (blue point). **f**, Optical microscope image of the device for demonstrating abrupt phase transition of the topological edge states. The abrupt phase transition point in **e** introduces an additional topological domain wall (blue dashed line) orthogonal to the original domain wall (black solid line) between regions I and II. **g**, Measured spatial distribution of elastic waves at the frequency of 48.35 MHz for the device in **f**. **h**, Mechanical amplitude spectra measured at points A–C in the device in **f**.

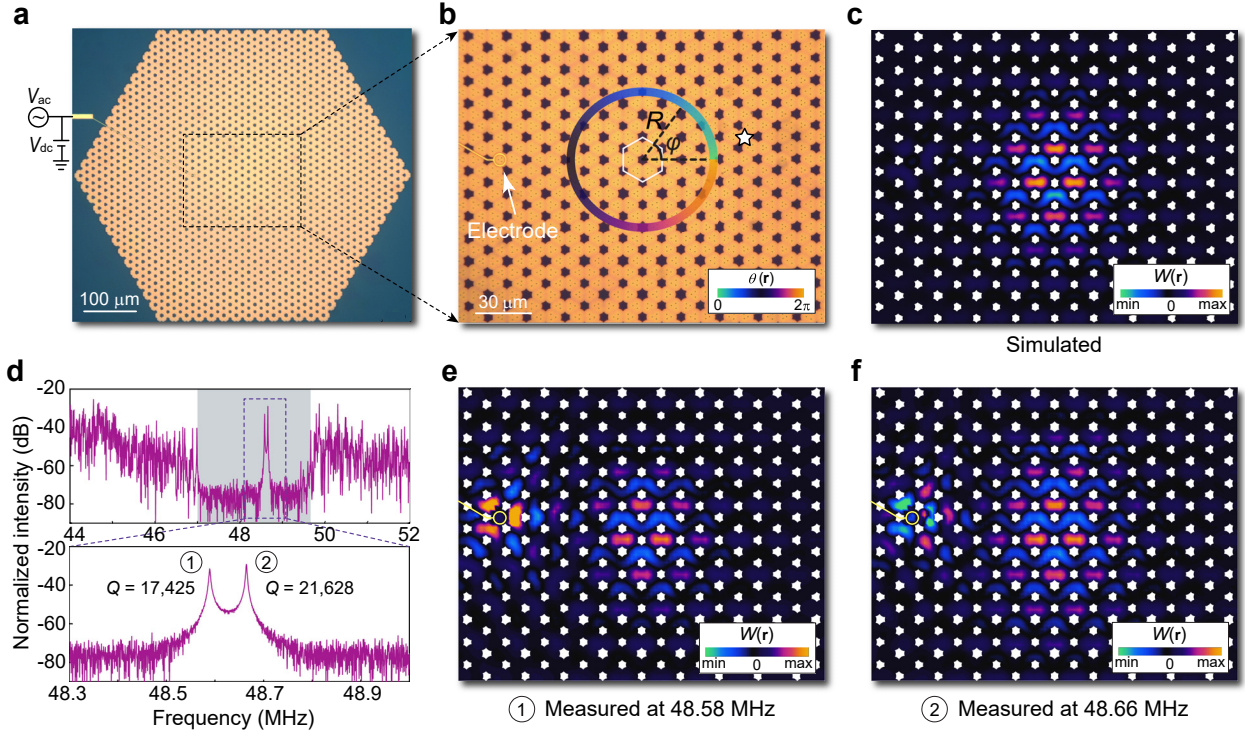

**Fig. 4 | Nanomechanical Dirac vortex states with an auxiliary orbital degree of freedom. a**, Optical microscope image of the device for demonstrating topological Dirac vortex state. **b**, Zoomed-in optical microscope image of the nanomechanical crystal. The white hexagon indicates the unit cell at the center of the vortex. The nanomechanical crystal has parameters $\delta_0(\mathbf{r}) = \delta_{max} \cdot \tan(R/R_0)$ and $\theta(\mathbf{r}) = \varphi$, both of which are a function of the position $\mathbf{r} = R \cdot (\cos\varphi, \sin\varphi)$. The electrode is near the center of the vortex for efficient actuation of the elastic waves. **c**, Simulated modal profile of the Dirac vortex state. **d**, Experimental amplitude spectra of elastic waves measured at the white star in **b**. **e**, **f**, Measured spatial distributions of elastic waves at the frequencies of 48.58 MHz (**e**) and 48.66 MHz (**f**).

15